\documentclass[twocolumn]{aastex63}  
\usepackage{graphics,epsf}
\usepackage{amsmath}                
\usepackage{amsfonts}               
\usepackage{amssymb}                
\usepackage{epsfig}                 
\usepackage{graphicx}
\usepackage{float}
\usepackage{color}
\usepackage[para,online,flushleft]{threeparttable}

\newcommand{\km}{{~\rm km}}
\newcommand{\s}{{~\rm s}}

\newcommand{\K}{{~\rm K}}
\newcommand{\erg}{{~\rm erg}}
\newcommand{\yr}{{~\rm yr}}

\newcommand{\AU}{{~\rm AU}}
\newcommand{\mum}{{~\rm \mu m}}

\begin{document}

\title{A rapidly fading star as a type II obscuring intermediate luminosity optical transient (ILOT) in a triple star system} 

\author[0000-0002-3592-1526]{Ealeal Bear}
\affiliation{Department of Physics, Technion, Haifa 3200003, Israel; \url{ealealbh@gmail.com}; \url{soker@physics.technion.ac.il}}

\author[0000-0003-0375-8987]{Noam Soker}
\affiliation{Department of Physics, Technion, Haifa 3200003, Israel; \url{ealealbh@gmail.com}; \url{soker@physics.technion.ac.il}}

\author[0000-0002-7840-0181]{Amit Kashi}
\affil{Department of Physics, Ariel University, Ariel, 4070000, Israel; \url{kashi@ariel.ac.il}}
\affil{Astrophysics Geophysics And Space Science Research Center (AGASS), Ariel University, Ariel, 4070000, Israel}


\begin{abstract}
We propose a triple-star scenario where the merger of two pre-main sequence low mass stars, ${\la 0.5 M_\odot}$, ejects a dusty equatorial outflow that obscures and temporarily causes the disappearance of a massive star, ${\ga 8 M_\odot}$. The merger of the low-mass inner binary powers a faint outburst, i.e., a faint intermediate luminosity optical transient (ILOT), but its main effect that can last for decades is to (almost) disappear the luminous massive star of the triple system. The typical orbital period of the triple system in about a year. The merger process proceeds as the more massive star of the two low-mass pre-main sequence star starts to transfer mass to the least massive star in the triple system and as a result of that expands. This \textit{type II obscuring ILOT} scenario in a triple star system might account for the fading, re-brightening, and then re-fading of the massive post-main sequence star M101-OC1.  It might recover in about 20-100 year.  Our study strengthens the claim that there are alternative scenarios to account for the (almost) disappearing of massive stars, removing the need for failed supernovae. In these scenarios the disappearing is temporary, months to decades, and therefore at later time the massive star explodes as a core collapse supernova even if it forms a black hole.  
\end{abstract}


\section{Introduction}
\label{sec:intro}

There are several different names to the heterogeneous group of transients with typical luminosities between those of classical novae and supernovae and their sub-classes (e.g. \citealt{Mouldetal1990, Bondetal2003, Rau2007, Ofek2008, Masonetal2010, Tylendaetal2013, Kasliwal2013, KashiSoker2016, Kaminskietal2018, Pastorelloetal2018, BoianGroh2019, Caietal2019, Jencsonetal2019, Kashietal2019, PastorelloMasonetal2019, Howittetal2020, Jones2020, Kaminskietal2020Nova1670, Andrewsetal2021, Blagorodnovaetal2021, Kaminskietal2021Nova1670, Klenckietal2021, Pastorelloetal2021a, Pastorelloetal2021b}). These names include intermediate-luminosity (red) transients, red novae, luminous red novae (e.g. \citealt{Jencsonetal2019, PastorelloFraser2019, Blagorodnovaetal2021}), and gap transients (e.g., \citealt{Kasliwal2011, Blagorodnovaetal2017, PastorelloFraser2019}). Processes that utilise gravitational energy to power such transients include mass transfer and stellar merger, including the onset of a common envelope evolution, (e.g., \citealt{SokerTylenda2003, Tylendaetal2011, Nandezetal2014, Kaminskietal2015b, Soker2016GEEI, SokerKashi2016TwoI, MacLeodetal2017, Gilkisetal2019, Segevetal2019, YalinewichMatzner2019, Schrderetal2020, MacLeodLoeb2020, SokerKaplan2021, Wadhwaetal2021}). 

As in our previous studies we will use the term intermediate luminosity optical transients (ILOTs; e.g., \citealt{Berger2009, KashiSoker2016, MuthukrishnaetalM2019}). We include under ILOTs all transients that are powered by gravitational energy even if they are not red and even if they are fainter than the typical luminosity of classical novae. In rare cases the powering of such an ILOT might be by a planet (e.g., \citealt{RetterMarom2003, Bearetal2011, KashiSoker2017a, Kashietal2019, Gurevichetal2022}), where in many cases the luminosity might be below those of classical novae.

This study deals with a non-eruptive star that suffers a rapid and very substantial fading. Nonetheless, we refer to the entire event of a variable star of this type as an ILOT because we propose a scenario where two stars (or even sub-stellar objects) merge (or collide) and shed dusty material that obscures the brightest star in the triple system. Namely, the scenario includes ingredients that we take from commonly studies of ILOTs.  
We present the scenario and the motivation to propose it in section \ref{sec:ILOTScenario}. In section \ref{sec:Evolution} we follow the evolution of the two merging stars to reveal their properties at the time of the fading of the massive star M101-OC1. We summarise in section \ref{sec:summary}. 

\section{The fading-ILOT triple star scenario}
\label{sec:ILOTScenario}
\subsection{Motivation}
\label{subsec:Motivation}

Our motivation to propose this scenario comes from the large fraction of triple stellar systems and from the possibility that there are no failed supernovae. 

Most papers in the last two decades attribute ILOTs to binary interactions of mass transfer, merger, and/or common envelope evolution (e.g., \citealt{Kashi2010, Kashietal2010, Tylendaetal2011, McleySoker2014, Nandezetal2014, Kaminskietal2015a, Kaminskietal2015b,  IvanovaNandez2016, Pejchaetal2016a, Pejchaetal2016b, Soker2016GEEI, Zhuetal2016, Blagorodnovaetal2017, MacLeodetal2017, MetzgerPejcha2017, MacLeodetal2018, Michaelisetal2018, HubovaPejcha2019, PastorelloMasonetal2019, Blagorodnovaetal2021}). As the fraction of triple star systems is not negligible (e.g., \citealt{MoeDiStefano2017}), we should also consider the possible outcome of these processes in triple star systems. We study one such a scenario here. 
 
\cite{Adamsetal2017} and \cite{Basingeretal2020} attributed the source N6946-BH1 in the galaxy NGC~6946 that experienced an outburst in 2009 March \citep{Gerkeetal2015} followed by a large luminosity drop to a failed core collapse supernova (CCSN) event. \cite{Humphreys2019} argue that the progenitor was a yellow hypergiant. 
Failed CCSNe that end the evolution of many very massive stars is a prediction of the neutrino driven CCSN explosion mechanism. According to this mechanism many massive stars with zero age main sequence (ZAMS) masses of $M_{\rm ZAMS} \ga 18 M_\odot$ do not explode (e.g., \citealt{Fryer1999, Ertletal2020, SukhboldAdams2020}). Most of the stellar mass collapses to form a black hole (BH), while the outer envelope is ejected at velocities much lower than those of typical CCSNe, leading to a transient event much fainter than typical CCSNe \citep{Nadezhin1980, LovegroveWoosley2013, IvanovFernandez2021, Tsuna2021PASJ}. 
In a recent study \cite{Neustadtetal2021} report the fading of M101-OC1, a blue supergiant that rapidly (almost) disappeared in the optical wavelengths. No outburst was observed and they found no dust. Before disappearing M101-OC1 had a rapid drop in luminosity, then it recovered, and only later its luminosity substantially dropped. \cite{Neustadtetal2021} consider this variability to be a failed CCSN. 

\cite{KashiSoker2017b} and \cite{Soker2021TypeII}, on the other hand, proposed and developed the type~II ILOT scenario for N6946-BH1 and similar events. According to the type~II ILOT scenario the binary interaction that powers the ILOT ejects a dense equatorial outflow of $ 0.1$--$1 M_\odot$. The outflowing equatorial dusty disk (torus) that the binary system ejects several years to several months before and during the outburst reduces the luminosity that an equatorial observer infers. The attenuation in wavelengths of  $\lambda< 5 \mum$ can be more than three orders of magnitude \citep{Soker2021TypeII}.
Another object that might be explained by this scenario is the massive star M51-DS1 that reappeared after its near-disappearance \cite{Jencsonetal2021J}.
This behaviour can be explained by a dust cloud that temporarily obscured the star, supporting the notion that dusty circumstellar matter can almost completely obscure massive stars.

Like in our earlier papers, the motivation to consider an alternative to the failed CCSN scenario is our view that there are no failed CCSNe. The alternative jittering jets explosion mechanism of CCSNe predicts that all stars explode, even when they form BHs (e.g., \citealt{GilkisSoker2014, GilkisSoker2015, Quataertetal2019}). The formation of a BH might end in a very energetic explosion, $E_{\rm exp} > 10^{52} \erg$, because the BH is likely to launch jets (e.g., \citealt{Gilkisetal2016Super}). 
Our earlier claim that there are no failed CCSNe has received a strong support from the recent observational study by \cite{ByrneFraser2022}. In a systematic search they find no transient that is consistent with a failed CCSN, and they put a strong limit on the fraction of possible failed CCSNe. 
    
We continue along the line of \cite{KashiSoker2017b} and \cite{Soker2021TypeII} and propose a type~II ILOT scenario 
for the almost disappearing in the optical of the star M101-OC1. To account for the observation of no pre-fading outburst we build a triple star type II ILOT scenario that includes ingredients of the type II ILOT scenario in binary stars. For having similar ingredients to other ILOTs and despite that there was no observation of an outburst we term this event \textit{type II obscuring ILOT}. 

\subsection{The scenario}
\label{subsec:TheScenario}
 
The scenario starts with three young stars in an hierarchical system that consists of a massive star (the primary) with an initial (ZAMS) mass of $M_{1,0} \ga 8 M_\odot$ and an inner binary (a short eccentric orbital period of $P_{\rm in} \simeq 1$--$4~{\rm weeks}$) of two low mass stars - the secondary $M_{2,0} \approx 0.2$--$1 M_\odot$ and the tertiary $M_{3,0} < M_{2,0}$. 
The orbital period of the outer binary, i.e., the inner binary of low-mass stars around the much more massive primary star, is $P_{\rm out} \simeq 0.3 \yr$ -- $3 \yr$, i.e., an orbital separation of about $a_{\rm out} \simeq 1$--$5 \AU$.  The most massive star might be beyond the main sequence, but because of its rapid evolution the age of the system is constrained to be $t_{\rm age} \la {\rm few} \times 10^7 \yr$.
The two stars of the inner binary (the secondary and the tertiary) merge because of a dynamical instability in this young system,  which might be caused by a fourth star in this system or by a passing star in the open cluster where this system was born. For example, the passing star perturbed the orbit of the close low-mass stars around the massive star, and this perturbation causes the triple star system to become unstable, leading to the merger.   

 In the scenario for M101-OC1 the merger takes place when the massive star has just left the main sequence. There is nothing particular in this time or in this line of sight of the observer that should be in or near the equatorial plane ( i.e., along the disk of the ejected mass). The merger might take place at any time during the early, $\la {\rm few}\times 10^7 \yr$, evolution of triple and higher order stellar systems. For example, V838~Mon is an ILOT where the commonly accepted model assumes that a merger of a massive and a low mass star took place in a young binary system (e.g., \citealt{TylendaSoker2006}). Some of these types of systems have the right inclination towards us to be obscured.

\begin{figure}[t]
	\centering
\includegraphics[trim=23.0cm 9.0cm 22.0cm 2.0cm ,clip, scale=0.50]{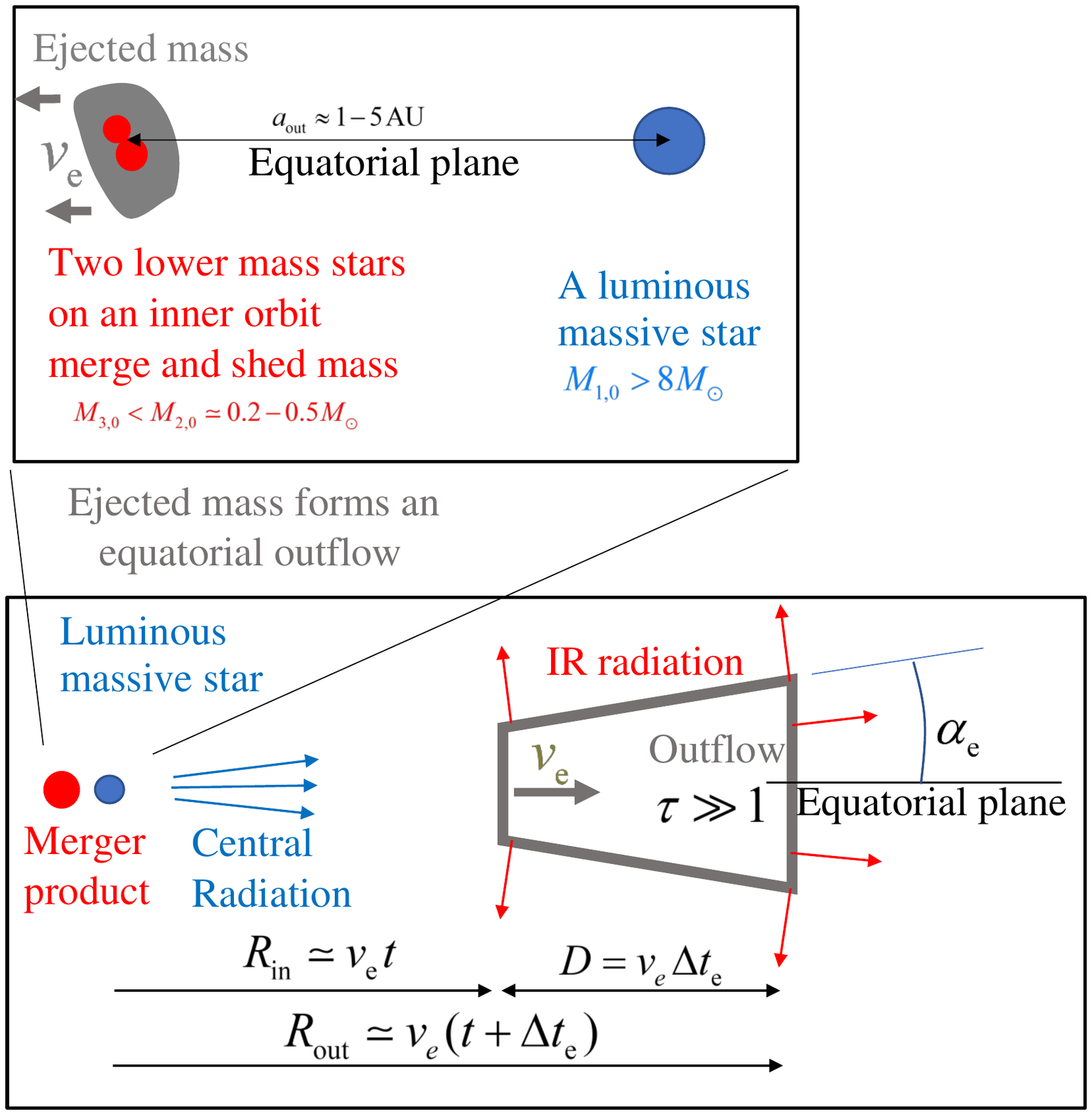}
\caption{A Schematic (not to scale) drawing of the triple type II obscuring ILOT scenario.
The upper panel depicts the merger of the two low mass stars (the secondary and the tertiary) that eject a dusty outflow in the triple equatorial plane. This occurs over a long enough time, about equal or larger than one orbital period, to eject mass in all equatorial directions. 
The lower panel contains a larger area of the meridional plane, i.e., the plane perpendicular to the orbital plane of the primary star and the merger product. We present the expanding equatorial outflow in only one half of the meridional plane, but the outflow is toroidal-like. Red arrows depict radiation that the dusty outflow reradiates. The $r$ coordinate is in the radial direction from the main radiation source (the primary) at the center, and the $z$ coordinate is perpendicular to the triple star orbital plane.
The inequality $\tau \gg 1$ refers to the optical depth inside the equatorial outflow in any direction (see also \citealt{Soker2021TypeII} for more details). }
	\label{fig:schecmatic}
\end{figure}

The merger process ejects a mass of $M_{\rm ej,e} \simeq 0.01$ -- $0.1 M_\odot$ in the equatorial plane of the triple system, as we present schematically in Fig. \ref{fig:schecmatic}. 
The merging inner binary inflates a large envelope due to the rapid mass transfer, and loses mass as it orbits the primary star. This process forms a disk of dusty ejecta along the orbit of the inner binary. Because of the radiation pressure from the primary star it slowly expands outward.   
According to the type~II obscuring ILOT scenario the ejected equatorial outflow substantially attenuates the radiation from the central source and shifts its emission to longer wavelength (red arrows in Fig. \ref{fig:schecmatic}). This explains the large fading in the visible and near~IR (wavelength of $\lambda \la 5 \mum$). 

To explain a non-eruptive fading event it is important that the primary star that accounts for most of the luminosity of the system does not participate in the merger process. The two lower mass stars did not reach yet their ZAMS phase, and they are somewhat inflated (section \ref{sec:Evolution}). The merger process does lead to an outburst, but at a lower luminosity than that of the stable luminosity of the massive star. 

We based the triple type II obscuring ILOT scenario on the type II ILOT scenario in binaries where an equatorial outflow obscures the central star after the eruption that the binary interaction induces  (\citealt{KashiSoker2017b, Soker2021TypeII}; see also \citealt{MacLeodetal2022}). The dusty equatorial outflow scatters and reradiates energy to all directions, and therefore reduces the emission towards an equatorial observer. This is different than a spherical shell that absorbs the radiation from the central source in all directions and, at equilibrium, reradiates the same luminosity to all directions. The equatorial dust outflow absorbs the radiation only from directions near the equatorial plane, and then reradiates to all directions, therefore emitting only a small fraction of the luminosity it absorbs to the equatorial directions. 
  
We follow \cite{Soker2021TypeII} and consider an equatorial outflow with velocity $v_e$ within an angle $\alpha_e$ on both sides of the equatorial plane that the stellar system ejected over a time period of $\Delta t_e$ ending at $t=0$, when the ILOT outburst takes place. In the present scenario there is no outburst, but this does not change the evaluation. Namely, the mass loss rate into the equatorial outflow started at time $t=-\Delta t_e$ and ended at $t=0$. We define the optical depth of the equatorial outflow to Thomson scattering in the radial direction, from inner to outer boundary, $\tau_{{\rm T},r}$, and its optical depth along the $z$ direction perpendicular to the orbital plane and including both sides of the equatorial plane $\tau_{{\rm T},z}$.
\cite{Soker2021TypeII} derived the ratio of the total (bolometric) luminosity that an equatorial observer would measure, $L_{\rm e, bol}$, relative to the luminosity they would have measured had there been no dusty equatorial outflow $L_{\rm I}$. We scale equation (13) from \cite{Soker2021TypeII} with typical values for the present scenario
\begin{eqnarray}
\begin{aligned}
\frac{L_{\rm e,bol}} {L_{\rm I}}
 &\approx 0.02
\left( \frac{\sin \alpha_e}{\sin 15^\circ} \right)^{5/2}
\left( \frac{\tau_{{\rm T},z}}{\tau_{{\rm T},r}} \right)^{3/2}
\\ & \times
\left( \frac {M_{\rm ej,e}}{0.1 M_2} \right)^{-1/2}
\left( \frac {M_2}{0.25 M_\odot} \right)^{-1/2}
\\ & \times
\left( \frac {v_{\rm e}}{100 \km \s^{-1}} \right)
\left(\frac{t+ \Delta t_{\rm e}}{11 \yr}\right)^{1/2}
\left(\frac{t}{10 \yr}\right)^{1/2}. 
\label{eq:Ltot}
\end{aligned}
\end{eqnarray}
This ratio holds for times longer than $t \ga a/v_{\rm e} \approx 1$--$2 \yr$. For shorter times we should  consider the source of the dust outflow that is located at $a \simeq 1$--$5 \AU$ from the massive luminous star.  
The luminosity $L_{\rm e,bol}$ that the equatorial observer infers is the bolometric value. Most of it is emitted in the IR, and the total luminosity for $\lambda<5\mum$ is less than 10\% of that value, i.e., $L_{\rm e}(\lambda<5\mum) < 0.1 L_{\rm e,bol}$ \citep{Soker2021TypeII}. 

\begin{figure}[t]
	\centering
\includegraphics[trim=23.0cm 14.0cm 22.0cm 4.8cm ,clip, scale=0.50]{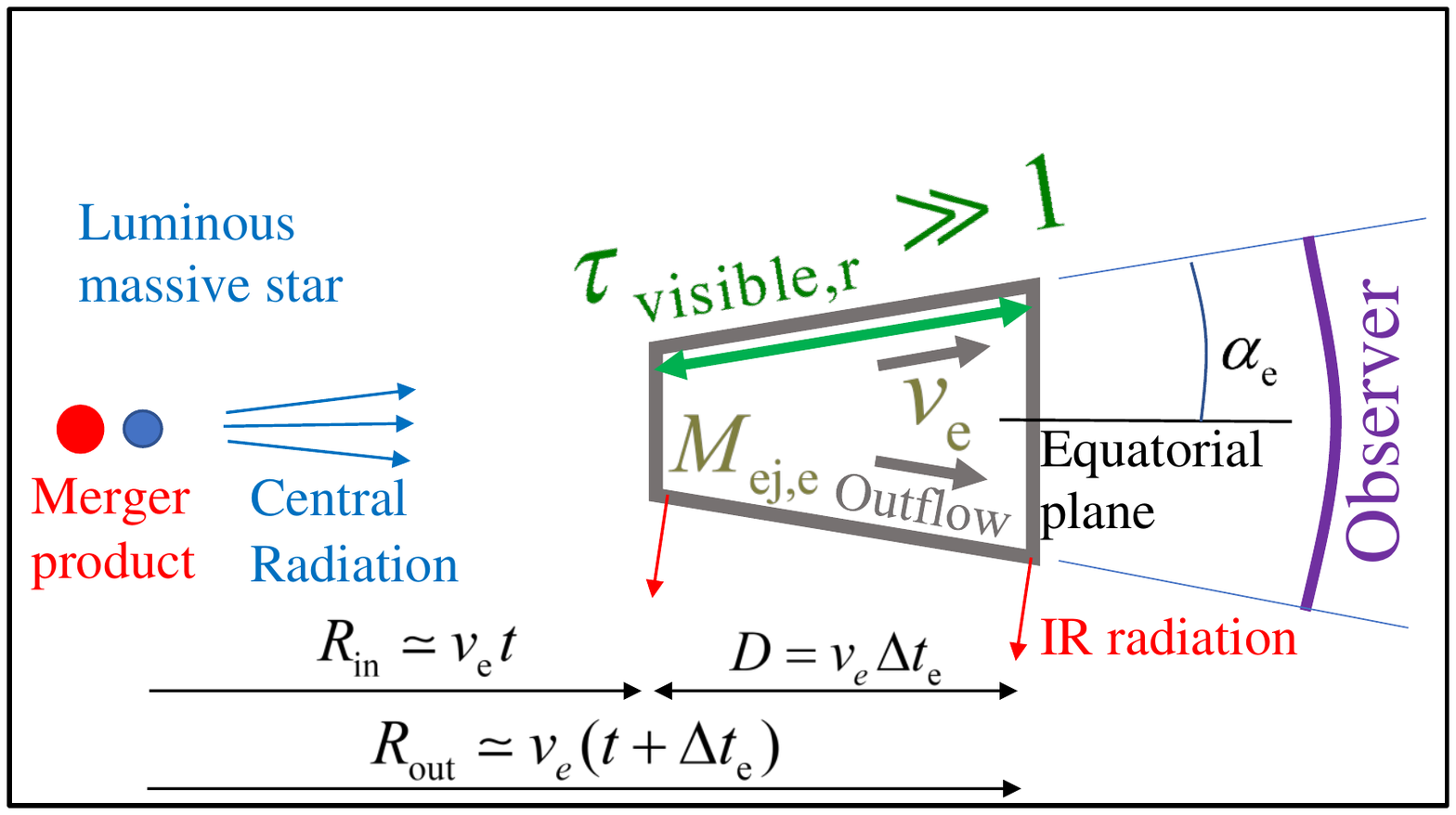}
\caption{A schematic (not to scale) drawing of the relevant quantities that enter into equation (\ref{eq:TauVr}). The double-head arrow is the path of the integration to calculate the optical depth in the visible from the central radiation source to the observer. On the right we show the possible directions for an observer in the scenario that we propose. }
	\label{fig:schecmatic2}
\end{figure}

Note that although we scale with the optical depth for Thompson scattering in equation (\ref{eq:Ltot}), because of the dust in the disk the actual optical depth is much larger than for Thomson scattering on electrons, e.g., by a factor of about 700, 200, and 40, in the visible band, at wavelength of  $\lambda = 1.25 \mum$ and at $\lambda = 9.7 \mum$, respectively (\citealt{Soker2021TypeII}). Namely, the dusty outflow is optically thick to the radiation from the hot star that supplies most of the radiation in the present scenario. 
  We can find the radial (through the disk in the radial direction) optical depth in the visible by scaling equation (10) from \cite{Soker2021TypeII} for the parameters of this study, and by taking the opacity to be 700 times as large as the Thomson one. This gives  
\begin{eqnarray}
\begin{aligned}
& \tau_{{\rm visible},r}  \simeq   300
\left( \frac{\alpha_{\rm e}}{15^{\circ}} \right)^{-1}
\left( \frac {M_{\rm ej,e}}{0.1 M_2} \right)
\left( \frac {M_2}{0.25 M_\odot} \right)
\\ & \times
\left( \frac {v_{\rm e}}{100 \km \s^{-1}} \right)^{-2}
\left(\frac{t+ \Delta t_{\rm e}}{11 \yr}\right)^{-1}
\left(\frac{t}{10 \yr}\right)^{-1}
 .
\label{eq:TauVr}
\end{aligned}
\end{eqnarray}
In Fig. \ref{fig:schecmatic2} we schematically draw the flow structure and the relevant quantities that enter into equation (\ref{eq:TauVr}). We emphasize the path along which the integration of the optical depth is performed, and the possible location for an observer in our proposed scenario.   

 From equation (\ref{eq:TauVr}) we learn that the disk can substantially attenuate the emission in the visible for decades for the above scaling. We can also deduce from equations (\ref{eq:Ltot}) and (\ref{eq:TauVr}) that the disk can substantially attenuate the emission in the visible for about 10 years after the ILOT event, even for an ejected mass ten times as low as what we use here. Namely, even an ejected mass of $M_{\rm ej,e} \simeq 0.002 M_\odot$ can cause a substantially attenuation in the visible for about a decade after the event. 

 \cite{Neustadtetal2021} observations of M101-OC1 show that the decline in the U and B bands are larger than that in the R band. This suggests dust obscuration. However, the moderate decline in the red (R band) also suggests that the optical depth in the red is not huge. It is possible that the ejecta mass is much lower than what we use for scaling here, even by as much as an order of magnitude. In that case M101-OC1  will return to normal in about 20 years.

 Overall, for the scaling of equations (\ref{eq:Ltot}) and (\ref{eq:TauVr}) with some uncertainty factor of few in the ejected mass $M_{\rm ej,c}$ and some uncertainty in the disk width $\alpha_{\rm c}$, we expect that the system M101-OC1 will return to its pre-2012 luminosity and color in about $20$--$100 \yr$ after outburst, i.e., in about the time period between 2035 and 2110.   

The minimum distance from the massive star where dust of sublimation 
temperature $T_{d,s}$ can survive is (e.g., \citealt{LaorDraine1993}) 
\begin{equation}
R_{\rm dust}  \approx  4 
\left(\frac {L_{\rm I}} {10^4 L_\odot} \right)^{1/2} 
\left(\frac {T_{\rm d,s}}{1500 \K} \right)^{-2.5} \AU ,  
\label{eq:Rdust}
\end{equation}
where ${L_{\rm I}}$ is the total luminosity of the central source. With an outflow velocity of $v_{\rm e} = 100 \km \s^{-1}$ the outflow reaches this distance in a time scale of two months. However, the merger of the two low mass stars takes place at $a \simeq 1-5 \AU$ from the luminous massive star, and so the outflow time to dust formation radius is only about a month or less.

We scale the duration of dusty equatorial ejection with one year because the variability of M101-OC1 lasted for about one year before it faded in 2014. The scaling of the time from that ejection is scaled with 10 years, as will be appropriate for the coming decade. The velocity of the equatorial ejecta is about the escape velocity from the triple system at about $a=2 \AU$.

\section{Stellar evolution}
\label{sec:Evolution}

 The aim of this section is to further emphasize two points. The first point to emphasize is that this system is young and the two low-mass stars of the close binary system are in their pre-main sequence phase. The more important goal is to demonstrate that once the mass transfer from the less dense star to the denser star in the pre-main sequence low-mass binary system starts, it can proceed and accelerate because the mass-losing star expands. 
 
We take the primary to be of mass $M_{\rm ZAMS}=12 M_\odot$ as this is about the mass that \cite{Neustadtetal2021} deduce for M101-OC1, when it has just left the main sequence at an age of $t=1.46 \times 10^7 \yr$, as we find in a simulation with \textsc{mesa} (see below). In the scenario we propose for the fading of M101-OC1 the source of the obscuring dust is the merger of two low mass stars, $M_{2,0} \approx 0.2$--$0.5 M_\odot$ and $M_{3,0} < M_{2,0}$, that are not yet on their respective ZAMS, and therefore they are larger that their respective sizes on the main sequence. These stars have luminosities that are about five orders of magnitude below that of the primary star. At the considered time the luminosity of the primary is $L_{\rm 1} = 2.2 \times 10^4 L_\odot$, while that of a secondary of mass $M_{2,0} =0.5 M_\odot$ it is $L_{\rm 2} = 0.1 L_\odot$. 

Consider then the merger of two low mass stars that takes place due to a perturbed eccentric orbit. The two stars might encounter a few grazing collisions before they merge (e.g., \citealt{Lombardietal1996, FreitagBenz2005}), as \cite{TylendaSoker2006} suggested was the case in the outburst of V838~Mon. To ensure a rapid merger we require that the mass losing star expands upon mass loss. For the low mass stars we study here, it is the secondary that is less dense, and hence it loses mass to the lower mass star. 

We therefore examine the evolution of pre-main sequence stars to rapid mass loss. In calculating the evolution of these stars we use the stellar evolution code \textsc{mesa} version 10398 \citep{Paxtonetal2011,Paxtonetal2013,Paxtonetal2015,Paxtonetal2018,Paxtonetal2019} with a metalicity of $Z=0.02$. To simulate the evolution of the low mass stars we follow the test suite example of \texttt{1M\_pre\_MS\_to\_wd}. 

\begin{figure}[t]
	\hspace{-2cm}
\includegraphics[trim=0cm 0.5cm 0cm 6.0cm ,clip, scale=0.6]{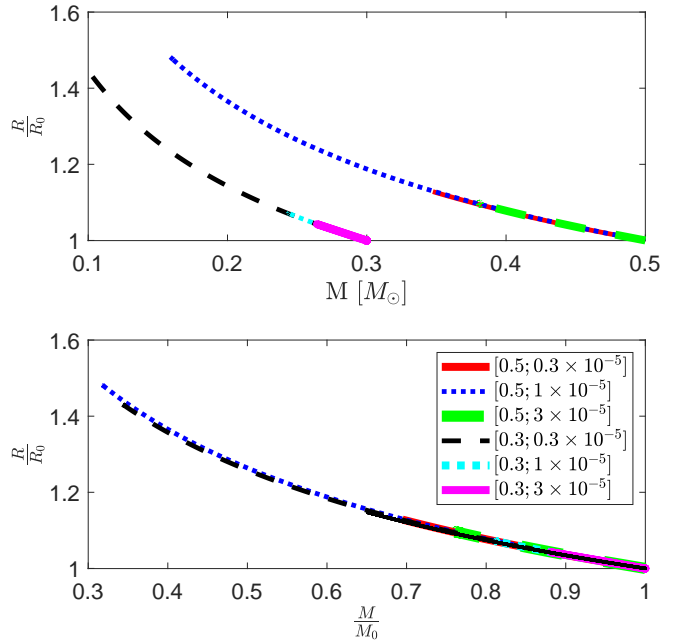}
	\vspace*{-4cm}
\caption{Upper panel: The ratio $R/R_0$ as function of the stellar mass, where $R(t)$ is the stellar radius and $R_0$ is the stellar radius at the time when we start to rapidly remove mass at $t=1.46\times 10^7\yr$. 
Lower panel: Similar to upper panel but the ratio $R/R_0$ is now a function of $M/M_0$. We preset results for two initial stellar masses $M_0$ and three different constant mass loss rates $\dot M$, as we show in the inset by $[M_{2,0}(M_\odot); \dot M (M_\odot \yr^{-1})$]. 
 Note that in these graphs the star evolves from lower right to upper left.  
} 
	\label{fig:RdivR0_vs_M}
\end{figure}

We expect the merger process to last for several weeks. However, due to numerical limitations we could not simulate mass removal on such short time scales. Instead, we simulate the mass loss process on much longer timescales of thousands to tens of thousands of years, but still much shorter than the thermal timescales of these stars. We present the results for two stellar models of the secondary star, which is the more massive one in the inner binary, in Fig. \ref{fig:RdivR0_vs_M}, where we also list the mass loss rates.

We perform the mass removal at an age of $t=1.46 \times 10^7 \yr$, just after the primary of $M_{\rm ZAMS,1}=12 M_\odot$ has left the main sequence. We find (Fig. \ref{fig:RdivR0_vs_M}) that the relative  expansions of the two stellar models are the same, and do not depend on the timescales in the range we simulate. The main conclusion is that the mass-losing star expands such that if the star starts to lose mass, either continuously or in a few encounters, the unstable mass transfer will continue to a merger. 

Key processes in our proposed scenario are that the merger product inflates a very large envelope, possibly $R_{\rm merger} \ga 0.5 \AU$, and ejects a mass of $M_{\rm ej,e} \approx  0.01 - 0.1 M_\odot$, similar to the ILOTs V838Mon (e.g., \citealt{Tylenda2005}) and V1309~Sco (e.g., \citealt{Tylendaetal2011, Nandezetal2014}). Such a large merger product might overflow its Roche lobe, and even overflows its second Lagrange point, hence losing mass in the equatorial plane of the triple star system. This mass loss process forms an equatorial slow outflow that might lead to the fading of the primary star for years to decades (equation \ref{eq:Ltot}). 

For an orbital period of $P_{\rm in} \simeq 1-4~{\rm weeks}$ of the inner low-mass binary  this process takes a few weeks to several months. Namely, the merging low-mass stars spread the dust that they eject during the merger process along their orbit around the primary massive star in a time period that is comparable to the outer orbital period, $P_{\rm out} \simeq 0.3 \yr -3 \yr$. At the first passage of the inner binary it ejects mass that might locally obscure the primary star, depending on the direction of the observer. This dimming might last for only a small fraction of the orbital period because the dust did not have enough time to spread along the entire orbit. Therefore, the primary star might temporarily return to its normal luminosity. This, we suggest, explains the recovering of M101-OC1 after its first dimming that lasted for $< 1\yr$ \citep{Neustadtetal2021}. 

The ILOT V838~Mon that reached a luminosity of $L_{\rm V838} \simeq 10^6 L_\odot$ that lasted for about two months (e.g., \citealt{Tylenda2005}), was most likely the outcome of a merger of $\simeq 6 M_\odot$ main sequence star with a pre-main sequence low mass companion of mass $\approx 0.3 M_\odot$ \citep{TylendaSoker2006}. The total energy that V838~Mon radiated was $\approx 5 \%$ of the gravitational energy that the merger process releases. The ILOT V1309~Sco that resulted from a lower-mass binary system had a luminosity of $\approx 3 \times 10^4 L_\odot$ for about a months, and then declined to $<3 \times 10^{3} L_\odot$ \citep{Tylendaetal2011}. 

In our proposed scenario for M101-OC1 and similar events the merger is of two stars of masses $M_{3,0} < M_{2,0} \simeq 0.2-0.5 M_\odot$, but $M_{2,0}$ is lower than the mass of the primary star in the binary system that merged to power V1309~Sco (e.g., \citealt{Nandezetal2014}). We expect that the peak luminosity of the merger of the two low mass stars in our proposed scenario be $<10^4 L_\odot$, which might be lower even in the visible. It is possible that the sparse observations of M101-OC1 missed the several weeks outburst between mid-2009 and the end of 2011. The giant merger product might be red and keep a luminosity of ${\rm several} \times 100 L_\odot$ for years, as its thermal time scale is hundreds of years.   
  
The star M101-OC1 did not disappear, but rather became dimmer and redder \citep{Neustadtetal2021}. We suggest that both the obscured primary massive star and the merger product of the two lower mass stars contribute to the residual luminosity of M101-OC1.

\section{Summary}
\label{sec:summary}

The main claims of the present study are that (1) there is an alternative to the failed CCSN scenario that might account for the (almost) disappearing of massive stars, and (2) triple star evolution can form an ILOT that causes fading even without a large outburst.
The first claim strengthens our earlier similar claim \citep{KashiSoker2017b, Soker2021TypeII} where we study type II ILOTs. The second one is an extension of the type II ILOT scenario to the presently proposed triple type II obscuring ILOT scenario (Fig. \ref{fig:schecmatic}).

Like in the type~II ILOT scenario, in our newly proposed scenario a dusty equatorial outflow causes the fading of the system (equation \ref{eq:Ltot}). But unlike the type~II ILOT scenario where the binary interaction that powers the ILOT also ejects the dusty outflow, in the triple type~II obscuring ILOT scenario an inner low-mass binary merges to supply the dusty outflow that obscures a massive and luminous star at a larger orbital orbit. The merger of the inner binary results in an outburst, but a relatively faint one. The expansion of the mass-losing star in the low-mass inner binary (Fig. \ref{fig:RdivR0_vs_M}) facilitates the merger process with mass ejection. 
We did not simulate the merger and triple-star dynamics. These processes will require more extended numerical studies.
  
We suggest (section \ref{sec:Evolution}) that the triple type II obscuring ILOT scenario might account for fading, re-brightening, and fading again of M101-OC1 as reported by \cite{Neustadtetal2021}. 

The question of whether the large fading or even disappearance in the visible band of massive ($M_{\rm ZAMS} \ga 10 M_\odot$) stars is due to a failed CCSN or to an ILOT (or ILOT-type) event where stellar merger ejects an equatorial dusty outflow of mass $M_{\rm ej,e} \approx 0.01-1 M_\odot$ (Type II ILOT) has far reaching implications to the explosion mechanism of CCSNe.
The neutrino driven explosion mechanism predicts that many massive stars can collapse to form a BH in a failed CCSN (section \ref{sec:intro}). This is the explanation that \cite{Adamsetal2017} and \cite{Basingeretal2020} adopted for the almost disappearance of N6946-BH1, and \cite{Neustadtetal2021} adopted for the almost disappearance of M101-OC1. \cite{KashiSoker2017b} and \cite{Soker2021TypeII}, on the other hand, adopted the jittering jets explosion mechanism according to which even when a core collapses to form a BH it launches jets to power a luminous CCSN (section \ref{sec:intro}; \citealt{ShishkinSoker2022, Soker2022} and references therein). They accounted for the almost disappearance of N6946-BH1 by the type II ILOT scenario in a binary system. 
The recent study by \cite{AntoniQuataert2022} strengthens the claim (e.g., \citealt{GilkisSoker2014}) that even if the core collapses to form a BH a bright transient event (i.e., a supernova) takes place. 

We end by predicting that massive stars that observationally disappear without explosion will reappear after a time scales of years to hundreds of years.  As for the event of M101-OC1, we estimate that it might be back to normal in the time period of
2035 to 2110. 

\acknowledgments

 We thank an anonymous referee for helpful comments.  
This research was supported by a grant from the Israel Science Foundation (769/20) and a grant from the Asher Space Research Fund at the Technion.
Part of the simulations in this paper were performed on the \textit{GALAXY} high performance computing cluster in Ariel University.
\newline
The data is available on Zenodo under an open-source 
Creative Commons Attribution license: 
\dataset[doi:10.5281/zenodo.6510095]{https://doi.org/10.5281/zenodo.6510095}.

\label{lastpage}
\end{document}